# Reactive Two-Step Additive Manufacturing of Ultra-high Temperature Carbide Ceramics

Adam B. Peters[1], Dajie Zhang[1,2], Dennis C. Nagle[1,2], James B. Spicer[1,2]*

[1]Department of Materials Science and Engineering, The Johns Hopkins University
3400 North Charles Street, Baltimore, MD 21218
[2]The Johns Hopkins Applied Physics Laboratory, Research and Exploratory Development
Department, 11100 Johns Hopkins Road, Laurel, MD 20723

**Graphical Abstract**

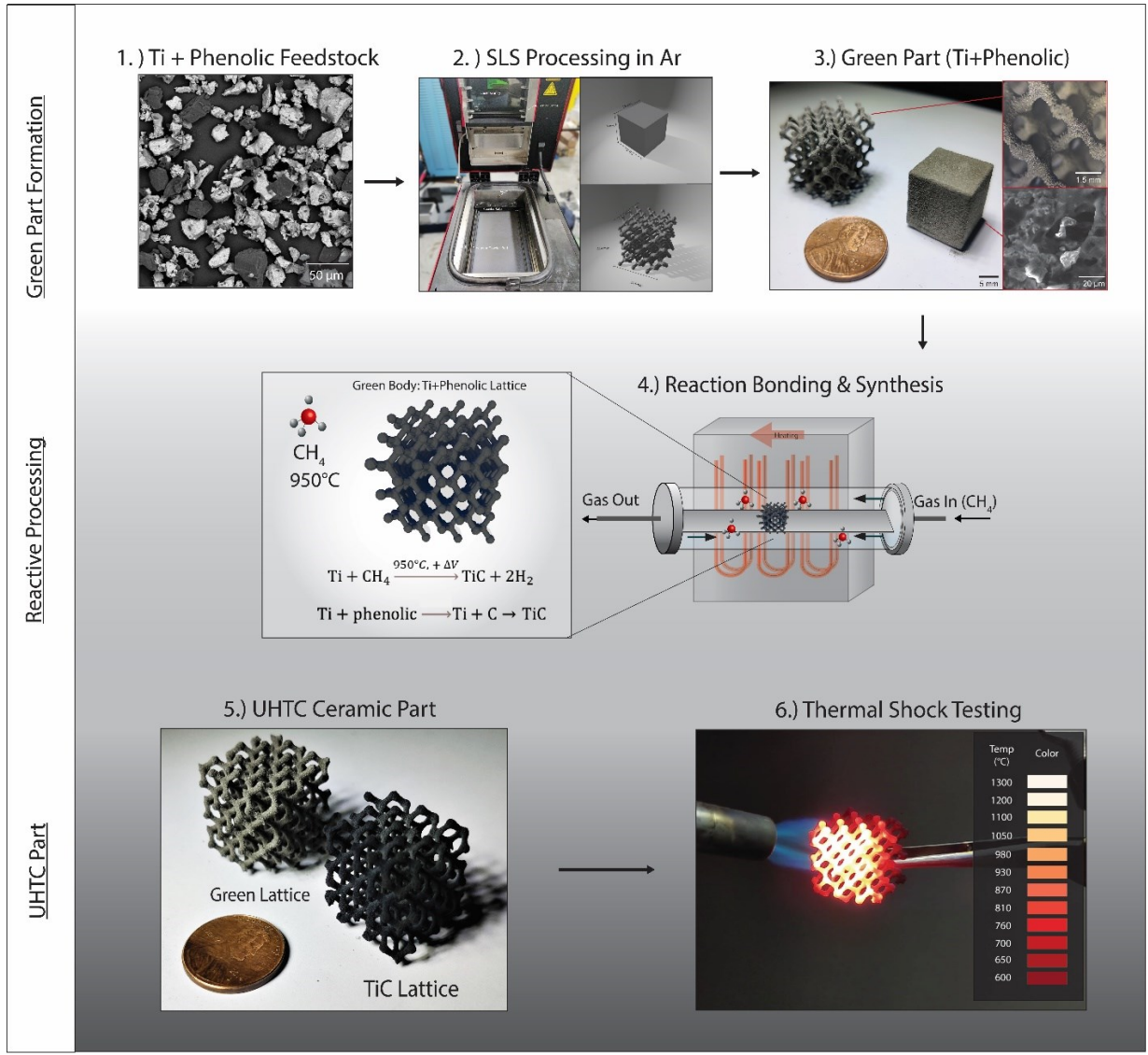

**Abstract**

Ultra-high-temperature ceramics (UHTCs) are candidate structural materials for applications that require resiliency to extreme temperatures (>2000°C), high mechanical loads, or aggressive oxidizing environments. Processing UHTC transition metal carbides as standalone materials using additive manufacturing (AM) methods has not been fully realized due to their extremely slow atomic diffusivities that impede sintering and large volume changes during indirect AM that can induce defect structures. In this work, a two-step, reactive AM approach was studied for the formation of the ultra-high temperature ceramic $TiC_x$. Readily available equipment including a polymer powder bed fusion AM machine and a traditional tube furnace were used to produce UHTC cubes and lattice structures with sub-millimeter resolution. This processing scheme incorporated, (1) selective laser sintering of a Ti precursor mixed with a phenolic binder for green body shaping, and (2) *ex-situ*, isothermal gas-solid conversion of the green body in $CH_4$ to form $TiC_x$ structures. Reactive post-processing in $CH_4$ resulted in up to 98.2 wt% $TiC_{0.90}$ product yield and a reduction in net-shrinkage during consolidation due to the volume expansion associated with the conversion of Ti to TiC. Results indicated that reaction bonding associated with the Gibbs free energy release associated with TiC formation produced interparticle adhesion at low furnace processing temperatures. The ability to bond highly refractory materials through this type of process resulted in structures that were crack-free and resisted fracture during thermal shock testing. Broadly, the additive manufacturing approach presented could be useful for the production of many UHTC carbides that might otherwise be incompatible with prevailing AM techniques that do not include reaction synthesis.

# 1. Introduction

Additive manufacturing (AM) is the formalized term for what is popularly known as 3D printing or rapid prototyping. The basic principle of AM is that 3-dimensional parts can be produced in a layer-by-layer fashion from a digitally generated model [1]. Over the last several decades, AM has become a highly attractive technique for the fabrication of complex and intricately shaped components [2]. AM of metals and polymers has progressed into a relatively mature technology, but this is not the case for refractory ceramic materials [3]. Non-oxide ceramics (carbides, nitrides, and borides) have highly desirable properties including high thermal and electrical conductivity as well as resilience to prolonged exposure to high-temperatures, chemically reactive conditions, radiation, stress, and mechanical wear. A subset of these non-oxides--known as ultra-high temperature ceramics (UHTCs)—have the highest melting points of any binary compounds (exceeding 3000 °C) as well as thermal and chemical stability in the air above 2000 °C. Due to their extreme refractory characteristics, interest in UHTC component fabrication has largely been motivated by the unmet materials requirements for aerospace, rocket propulsion, and hypersonic thermal protection systems [4]. UHTC carbides including HfC, ZrC, TaC, and TiC have received attention for hypersonic applications such as thermal protection systems, nozzle throats and radomes that operate under conditions of high heat flux, aggressive oxidizing environments, and rapid heating/cooling rates during flights [4]–[8].

The production of complex UHTC geometries using additive manufacturing or traditional ceramics processing techniques is both difficult and costly. Strong covalent-ionic and metallic bonds in these materials inhibit sufficient atomic mobility to relieve thermally-induced stresses during additive processes and can lead to decomposition when heated to temperatures that produce mobility [9]. Both traditional dry powder and colloidal shaping techniques are very difficult to use and, consequently, high post-processing temperatures and pressure-assisted techniques are needed to produce dense components. These methods often limit geometric complexity to simple axially-symmetric shapes (like cylinders or tiles) or components without internal features [8]. When refractory ceramics are formed using AM techniques, high-temperature consolidation (sintering) of granular materials require a binder phase or organic additives (dispersants, binders, plasticizers, lubricants, etc.) to provide the desired rheological and cohesive properties on non-reactive feedstocks. For AM of UHTC materials, slow atomic diffusion hinders consolidation and sintering of non-oxide particles: high temperatures (in excess of 2000 °C), slow heating rates (0.1-2°C/hr) and hot isostatic pressing are used to prevent defects and achieve the desired mechanical integrity [10], [11].

In this work, a commercial polymer powder bed fusion machine was used for the investigation of an AM processing method that incorporates indirect selective laser sintering of metal precursor materials and conversion to the desired UHTC ceramic during post-processing. Using this process, the chemical conversion and volume changes associated with the production of geometrically complex titanium carbide (TiC) shapes were studied. TiC is an ultra-high temperature material with unique properties: high melting point (3067 °C), high hardness (2800 HV-the most of any carbide), extreme compressive strength (highest of any known material at 36,000 psi [12]), resistance to chemical attack, low coefficients of friction, and high electrical and thermal conductivity [13], [14]. TiC was selected as a model system representative of transition metal UHTCs (ZrC, HfC, TaC) that might also be produced using this method. Titanium carbides have been fabricated by electron beam or selective laser melting but are used as a low phase fraction constituent and reinforcing agent for metal-matrix composites (<35% wt%) [15], [16]. However, AM of TiC as a single-phase, standalone material has not been reported. Other non-reactive, multi-step AM strategies have been examined to produce non-oxide refractory ceramics but they have a range of limitations. For example, SiC parts have been formed through the shaping of SiC + binder then post-processed for densification where Si melt infiltration was required [17], [18]; Leu investigated the indirect selective laser sintering of $ZrB_2$ where post-processing at 2050 °C resulted in 54% shrinkage with 87% final density [10]; laser sintering was used to melt Si-containing C, then converted to ~93 wt% SiC during post-processing in inert atmospheres [19].

Rather than relying on non-reactive, thermally-driven sintering during high-temperature-post processing or direct laser (or electron beam) melting, the synthesis technique described here incorporates gas-solid reactions for conversion of reactive green body precursor materials to the desired ceramic. This method is analogous to and inspired by the formation of reaction-bonded silicon carbide and silicon nitride using traditional ceramics processing. Specifically, the reaction synthesis approach studied in this work can be summarized by following steps:

1. Inert selective laser sintering of Ti precursor containing an expendable, low melting temperature binder phase (phenolic resin) that is used to consolidate and shape a green body;
2. High-temperature, isothermal post-processing of preceramic sructures that react with $CH_4$ and binder-decomposition-products for UHTC carbide synthesis.

In this approach, two carbidization reactions lead to the formation of TiC:

$$\text{Ti} + \text{CH}_4 \xrightarrow{950°C,\ +\Delta V} \text{TiC} + 2\text{H}_2 \qquad \text{Eq. 1}$$

$$\text{Ti} + \text{phenolic} \rightarrow \text{Ti} + \text{C} \rightarrow \text{TiC} \qquad \text{Eq. 2}$$

The SLS/reaction synthesis approach investigated here was designed to (1) mitigate shrinkage that is typically associated with ceramics post-processing using the volume expansion of Ti to $TiC_{1.0}$ (~14.2 vol%); and (2) facilitate interatomic mobility for particle adhesion by leveraging large $\Delta G_r$ released during product formation. Chemical reactions can facilitate atomic mobility that lead to interparticle bonding in materials systems that are otherwise non-sinterable. For a chemical reaction, $\Delta G° \approx 20,000$ J/mol or more, a value significantly greater than the driving forces associated with applied stress or surface area changes alone [20]. For non-oxide materials and UHTCs having diffusion coefficients significantly lower than for many refractory oxides, it is very desirable to use this reaction energy to drive interparticle adhesion [21]. The use of reaction synthesis AM to standalone UHTC compositions might be used to construct complex refractory components for carbide catalysis, thermal protection systems, rocket propulsion, or other extreme environment applications [4], [22]–[28].

## 2. Methods

### 2.1 Experimental: SLS of Reactive Precursors using Commercial Powder Bed Fusion Followed by Reactive Processing.

For SLS processing, a Sinterit Lisa Pro® laser sintering machine was used. The SLS machine features a 5 W 808nm diode laser, X-Y accuracy ≤ 50μm, heated build platform, and maximum print size of 110mm × 160 mm × 230 mm for high-temperature (polymer) materials. The sealed build chamber was modified for compatibility with Ar gas and equipped with a dynamic $O_2$ monitoring device to prevent Ti oxidation during SLS green body shaping. The experimental configuration of the Lisa Pro® and an as-deposited powder bed is shown in Figure 1.

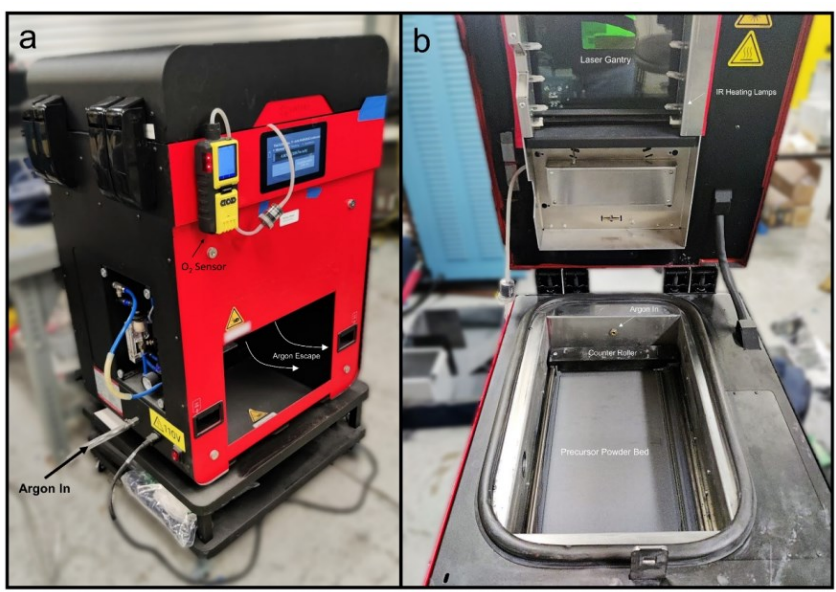

**Figure 1.** (a) Sinterit Lisa Pro® selective laser sintering machine modified for compatibility with variable argon flow. (b) Photograph of the interior of the build chamber containing precursor powder.

### 2.2 Print Geometries and Precursor Rationale

Two print geometries were selected for component fabrication: a 1.5 cm x 1.5 cm x 1.5 cm cube (to assess the influence of anisotropic volume changes, part density, and $CH_4$ penetration) and a diamond cubic lattice structure (to evaluate the spatial resolution and precision of the AM processing scheme). Digital illustrations used for printing the target test structures are shown in Figure 2. In subsequent studies, other shapes such as bend bars or dog bone tensile/compression test bars will be fabricated for mechanical testing.

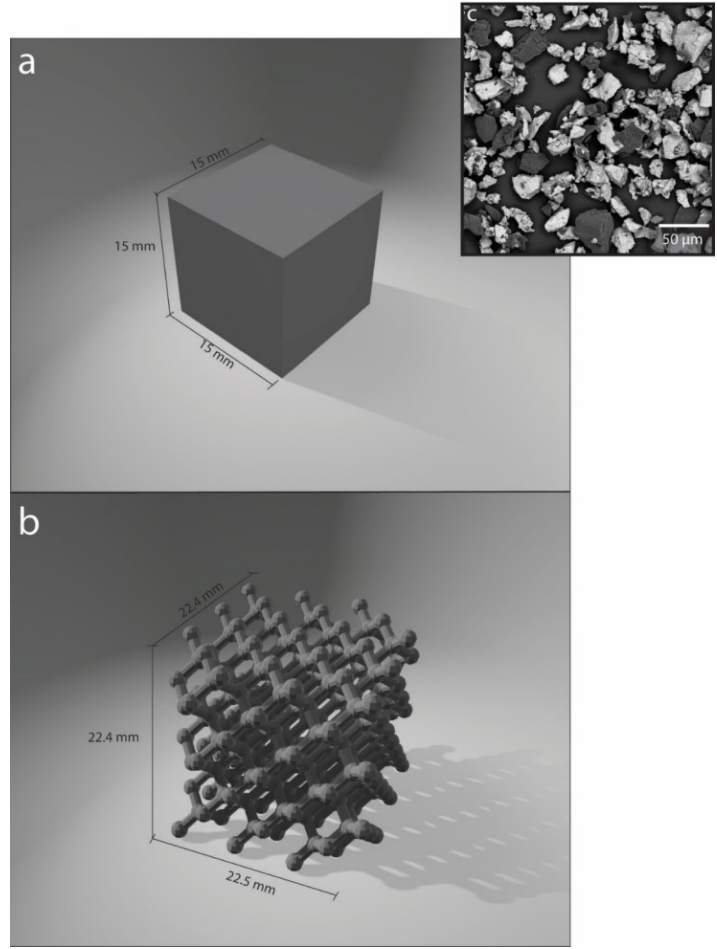

**Figure 2.** Digital representation of the printed test structures: (a) 15 mm x 15 mm x 15 mm cube and (b) diamond lattice structure. (c) shows a BSE-SEM micrograph of the 75/25 vol% Ti/phenolic precursor particle morphology, where large bright particles are Ti, and dark particles are phenolic.

The optical power output of the 5 W laser was maximized in the Sinterit Studio software. The scan speeds of the Sinterit Lisa Pro® were fixed and limited to 100 mm/s. Preliminary trials using Ar processing indicated that the average energy density was too low for direct sintering of Ti particles. Additionally, because of the safety risks associated with reactive laser processing in $CH_4$ without a discrete gas exhaust line in the Lisa Pro®, strategies employing *in-situ* gas-solid reactivity using $CH_4$ (akin to [29], [30], and [31]) were not employed. Rather, this indirect processing using $CH_4$ conversion of green body parts was used.

### 2.3 Ti+Phenolic Resin Precursor Formulation and SLS Processing

Ti powder (Atlantic Equipment Engineers Ti-107) and phenolic novolac resin (Hexion Durite AD-5614) were selected as the feedstock material for laser sintering in Ar and green body shaping. Good flowability is required for powder bed AM processes in which a counter roller is used to deposit thin layers of material. Relatively large particles (10-100 μm) enhance flowability and result in powder-packed densities between 25-45% [32]. Both the Ti powder and phenolic resin were chosen due to their <74 μm particle size and morphology which enabled reliable screening over the build platform.

Durite AD-5614 phenolic, in particular, was selected due to its robust bonding characteristics when cured, high carbon yield (58 wt%), and optimal decomposition temperature (950 °C) [33]. A BSE-SEM image showing the mixture of Ti and phenolic particles (75 vol% Ti, 25 vol% phenolics) is presented in Figure 2c. Since the number of backscattered electrons is proportional to the mean atomic number of the elemental components, bright particles in BSE images are associated with titanium due to its higher average atomic number [34], [35]. All precursor mixtures were mixed for 2 hrs in a roller mixer containing ceramic mixing media to ensure homogeneous particle distribution.

Only Ti was utilized in the feedstock (rather than a Ti/$TiO_2$ composite precursor), so volume expansion upon conversion to TiC (+14.2 vol% for Ti →$TiC_{1.0}$ [30], [36], [37] ) might largely compensate for consumption of the binder during pyrolysis. Initial conversion trials using ~14.2 vol% phenolic were conducted to test the lower limit for binder content. For shaping of the green body via SLS, the internal build chamber was set to 50 °C to help reduce residual stresses and pre-heat the phenolic so laser energy could efficiently bring the precursor mixture to the phenolic glass transition temperature. The melting/glass transition temperature of the Durite powder was estimated to be approximately 125 °C with curing temperatures occurring at 150 °C (taking roughly 60 seconds). Oxygen gas levels were dynamically monitored and reduced to <0.2 vol% $O_2$ before selective laser sintering. A summary of the processing parameters is given in Table 1.

**Table 1. SLS Laser Processing Parameters**

| | |
|---|---|
| Materials | 75 vol% Ti + 25 vol% Phenolic |
| SLS Processing Gas | 100 vol.% Ar (8L/min) |
| Target Product | $TiC_{1.0}$ |
| Wavelength | ($\lambda$) = 808 nm |
| Average Power | (P) = 5.0 W |
| Scan Speed | (V) = 100mm/sec |
| Resolution | (R) = ~50 μm |
| Deposition Layer Thickness | ($D_{layer}$)=175 μm |
| Powder Bed Temperature | ($T_{Bed}$)=50 °C |

Typical binder phases (polyamides, amorphous polystyrene, and polypropylene) used for indirect selective laser processing of ceramics can constitute ~50-70 vol% of the feedstock [38]. Preliminary tests incorporating 14 vol% phenolic binder produced particles that were very weakly bound in the green body, leaving the shape with similar mechanical characteristics to those of damp sand. This made handling the laser-sintered body impractical and small features prone to damage upon removal from the powder bed, and this is shown by photographs in Figure 3.

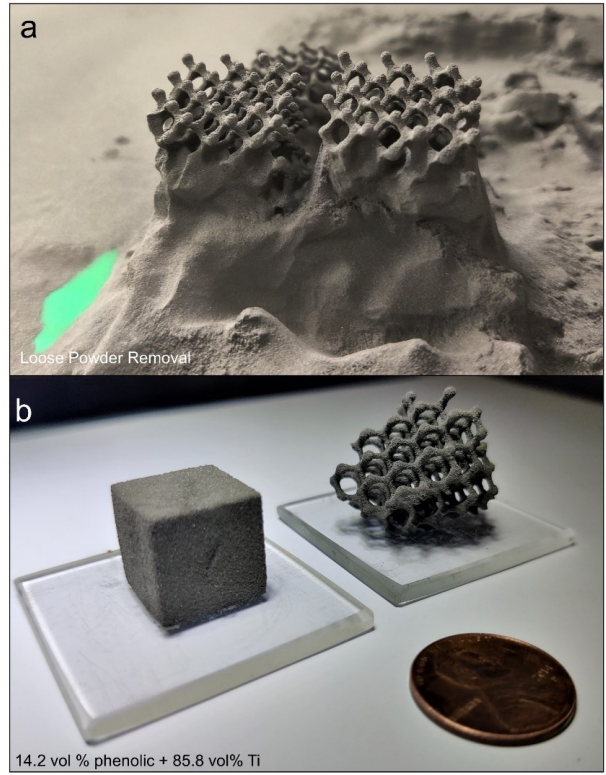

**Figure 3.** Green components formed from 14.2 vol% phenolic resin powder + 85.8 vol% Ti during removal from the powder bed (a) and after loose powder was removed (b). The structures shown in (b) had low interparticle binding leading to damaged features and disintegration of the cube and lattice corners when handled.

To increase the mechanical properties of the green part, the phenolic resin content was increased to 75 vol% Ti powder + 25 vol% phenolic resin powder, and this composition was found to be a reliable precursor formulation for ease of handling and robustness. The final composition and characteristics of the precursor material used for two-step TiC AM and reaction synthesis are presented in Table 2.

**Table 2. Characteristics of Selected Precursor Materials Used for Indirect TiC Formation**

|           | Manuf./Name           | Feedstock Volume | Particle Size | Density (g/cm$^3$) | Carbon Yield      | Carbonization Temp |
|-----------|-----------------------|------------------|---------------|--------------------|-------------------|--------------------|
| Ti Powder | AEE Ti-107            | 75%              | ≤74 μm        | 4.506              | -                 | -                  |
| Phenolic  | Hexion Durite AD-5614 | 25%              | ≤74 μm        | ~1.1               | ~58 wt% [33]      | 950°C              |

While the volume loss related to phenolic pyrolysis might not be entirely compensated by Ti carburization, this study serves to broadly investigate the reaction synthesis methods for AM of UHTCs. Compared to the 50-70 vol% of binder materials typically used in ceramic feedstocks, a reduction ~25% reduction in binder volume is a significant improvement that might mitigate excessive shrinkage during post-processing.

## 2.4 Tube Furnace Post-Processing

Using the composition in Table 3, pyrolysis of the phenolic binder phase could generate enough carbon for 31.3% conversion to stoichiometric TiC$_{1.0}$. Gas-solid processing in CH$_4$ was required to complete the reactivity of the green part to TiC. After SLS, the structures were post-processed in 80/20 vol% Ar/CH$_4$ using the tube furnace apparatus described previously [36], [37]. An alumina tube rather than a quartz tube was used to permit higher processing temperatures of up to 1350 °C.

Three conversion regimes using two different heating schedules in either inert or reactive gas were used for conversion and consolidation. Variations in heating for different processing atmospheres were used to assess the influence of conversion on volume change and carbide yield. In all cases, an initial dwell time of 0.5 hrs at 160 °C was used to cure the binder phase and fix the part shape before ramping to peak temperatures. The ramp-up and ramp-down rates after phenolic cross-linking (160 °C) were fixed at 100 °C/hr. After curing, the temperature was increased either to 950 °C (for gas-solid conversion, then sintering at 1350 °C) or directly to 1350 °C (for pre-sintering, followed by reaction at 950 °C). Figure 4 and Table 3 summarize the different post-processing reaction schemes.

**Table 3. Summary of Post-processing Regimes**

|                            | Scheme I: Inert Processing (Control) | Scheme II: React, Post-Sinter | Scheme III: Pre-Sinter, React |
|----------------------------|--------------------------------------|-------------------------------|-------------------------------|
| Ramp Up Rate (above 160°C) | 100°C/hr                             | 100°C/hr                      | 100°C/hr                      |
| Ramp Down Rate             | 100°C/hr                             | 100°C/hr                      | 100°C/hr                      |
| Dwell 1                    | 950°C (Ar)                           | 950°C (CH$_4$)                | 1350°C (Ar)                   |
| Dwell 2                    | 1350°C (Ar)                          | 1350°C (Ar)                   | 950°C (CH$_4$)                |

The rationale behind each set of reaction conditions can be summarized as follows:

- Scheme I is a control process. Samples are heated to 950 °C (the phenolic decomposition and CH$_4$ reaction temperature) in an inert atmosphere. This is done to estimate the TiC$_x$ yield from C$_{(s)}$ supplied by phenolic decomposition without gas-solid conversion. After heating and reactive dwell at 950 °C, samples were sintered and consolidated at 1350 °C.
- Scheme II utilizes the same heating schedule as Scheme I but incorporates CH$_4$ gas-solid reactions to convert unreacted Ti to TiC during the 950 °C dwell. After gas-solid conversion, the sample was sintered at 1350 °C to aid consolidation.
- Scheme III incorporates both C$_{(s)}$ and CH$_4$ reactivity, however, the sample was first sintered at 1350 °C to understand how the brown part density (the green part after sintering) affects CH$_4$ gas-penetration and conversion to low stoichiometry TiC$_x$ along with the corresponding impact on reaction bonding and part volume change.

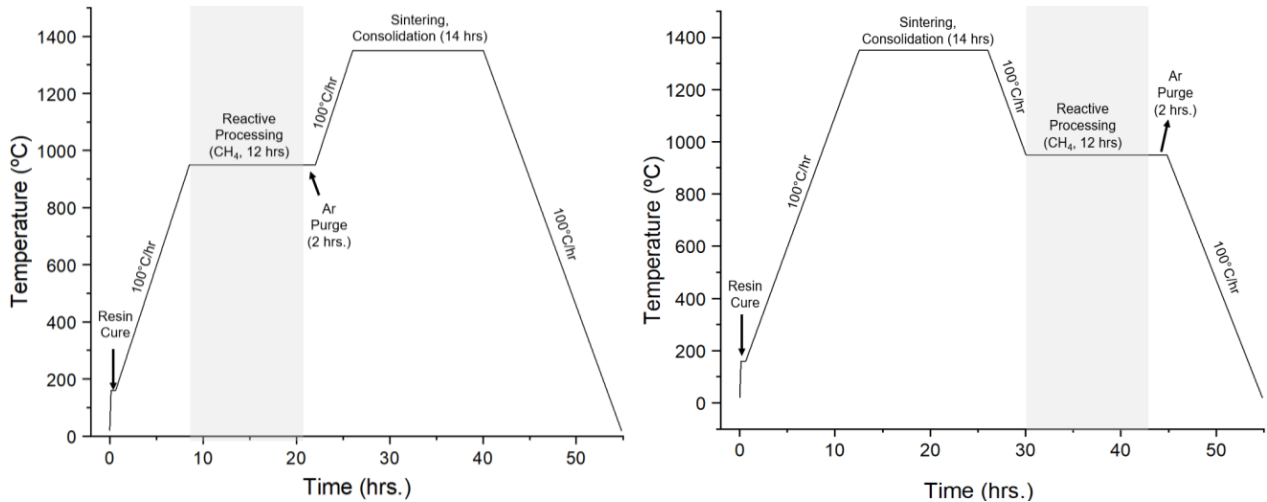

**Figure 5.** Heating schedules for reaction conversion of 75 vol% Ti + 25 vol% phenolic green state parts to TiC components.

The time for each segment of the heating schedule was identical. Margiotta and Trammell noted that carbonization of Durite 5614 phenolic occurs efficiently between 950-1000 °C [33], [39]. A reaction temperature of 950 °C for gas-solid reaction with CH$_4$ was used to mitigate carbon deposition that was observed at higher temperatures. The total flow rate was maintained at 250 standard cubic centimeters per minute (SCCM) during heating in inert (100 vol% Ar) or reactive (80/20 vol% Ar/CH$_4$) atmospheres. For samples that were processed in reactive atmospheres, CH$_4$ was introduced into the furnace at the 950 °C dwell temperature for a dwell time of 14 hrs. The introduction of CH$_4$ at the peak dwell temperature was selected in an effort to maximize the $\Delta G_r$ and facilitate reaction bonding between particles without spontaneous gas-phase CH$_4$ decomposition and carbon deposition that might reduce open porosity and inhibit conversion. Due to the slow decomposition of phenolic and slow solid-state carbon reactivity for carbide formation compared to gas-activated conversion, CH$_4$ is likely the dominant carbon source for the conversion process [40].

SLS processed and converted materials were characterized using x-ray diffraction (XRD) to determine the rate of conversion to TiC$_x$. Quantitative phase characterization was performed from 20° to 80° 2θ using Rietveld refinement with Match! software. XRD was conducted on the cube sample surfaces and on cross-sections. Surface characterization provided phase composition data when gas-solid reactivity was not limited by CH$_4$ diffusion through the inter-particle matrix; XRD of the cross-section was used to estimate the average conversion achieved through the ~15 mm sample thickness. A combination of optical and SEM microscopy methods were used to characterize the sample microstructures using procedures described previously [29], [36], [37].

The density of fully dense 75 vol% Ti/25 vol% phenolic was estimated to be ≈ 3.65 g/cm$^3$ (assuming the trace phenolic decomposition during SLS processing in Figure 7). To estimate the density of as-printed green cubes, this density was compared to the mass of the as-printed green cube over its measured volume. The estimated volume fractions of solid phases in Table 5 were similarly computed by calculating the theoretical density of a fully dense as-converted compositions (using a weighted fraction of product densities obtained from XRD/Rietveld refinement) and then comparing the fully dense value to the mass over the boundary volume of the as-converted component.

## 3. Results

### 3.1 Recovery and Morphology of Green Parts

SLS-processed, titanium green bodies were removed from the build chamber, handled, and separated from loose particles without notable damage to either the cube or the fine lattice structures. Once removed from the build plate they were inspected using optical microscopy. The morphology of the Ti/phenolic lattice and cube are shown in Figure 6.

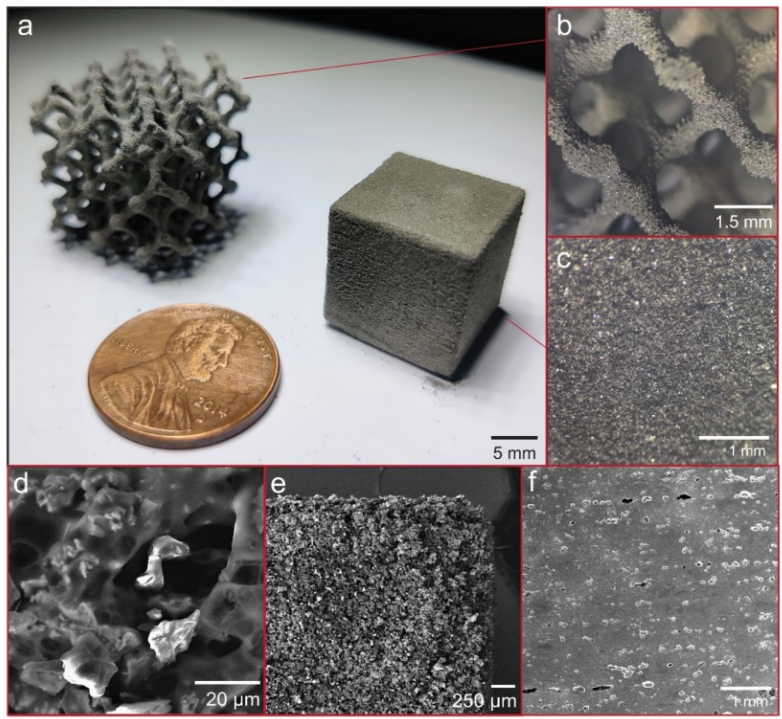

**Figure 6.** Photographs of the 75 vol% Ti powder+ 25 vol% phenolic resin (92.5/7.5 wt%) SLS processed into the diamond lattice and cube structures (a). Photomicrographs in (b,) (c), and (e) show the surface roughness and resolution of the printed structures. (d) depicts the Ti particles bound in melted phenolic after SLS processing, and (f) depicts a polished cross-section of the epoxy-impregnated green part. Bubbles in the cross-section may partly be the result of gasses trapped from the rapid resolidification of phenolic during green part SLS.

The average, as-printed dimensional deviations from the specified 15 mm x 15 mm x 15 mm cube were 0.0%, -0.7%, and +2.7% in the x, y, and z directions respectively for five samples. The deviation for the z-direction is likely due to the selection of the layer deposition height parameter (175 µm) and the rough Ti particle morphology that did not optimally pack. These deviations can be reduced using the AM software. The unreacted, as-printed density of the green bodies was determined to be ~31.8% dense. This value falls within the 25-45% range reported previously [32]. In future iterations, a significant increase in green part density might be achieved through the optimization of particle packing using spherical particles, a bi-modal distribution of particles, or alternative slurry-like deposition approaches [32].

### 3.2 Conversion Propensity and Phase Analysis Following Post-Processing in CH$_4$

The XRD spectra of the unreacted precursor materials and the green-state sample are shown in Figure 7.

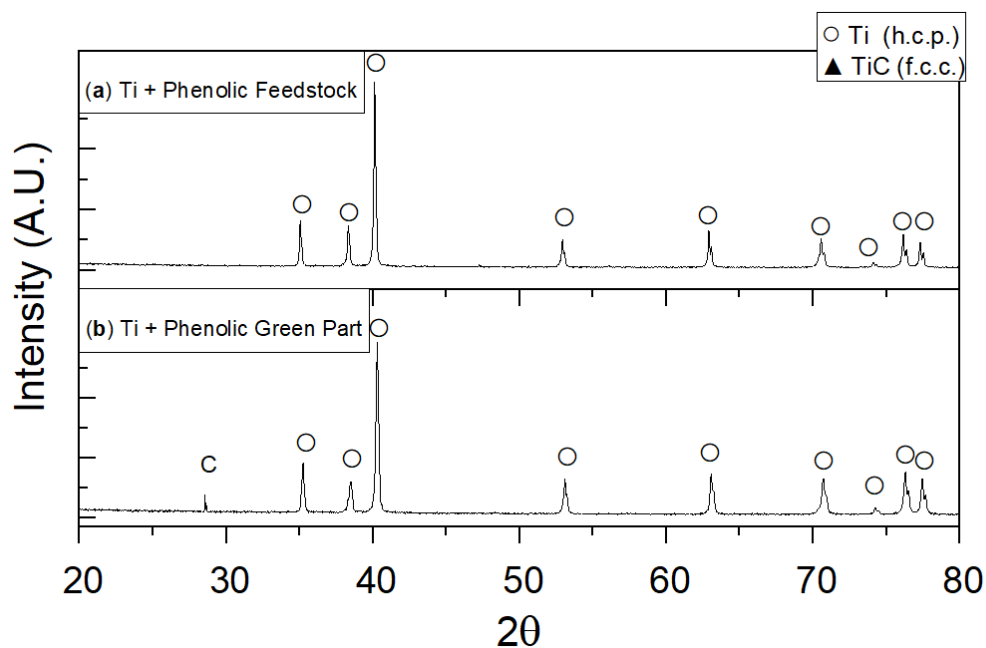

**Figure 7.** XRD spectra of unreacted 75 vol% Ti powder+ 25 vol% phenolic resin feedstock (a), green-state sample (b) after SLS processing to structures shown in Figure 6.

XRD characterization of Ti+phenolic precursor in Figure 7a indicates primary peaks associated with α-Ti while the amorphous structure of the phenolic resin did not result in a well-defined diffraction pattern. Phenolic resins typically display broad amorphous humps from 5-25 degrees 2θ. SLS processing of the 75 vol% Ti powder+ 25 vol% phenolic did induce partial decomposition of the phenolic binder (as indicated by the C peak at 28 degrees 2θ), but not *in-situ* carbide formation (Figure 7b). Therefore, conversion to TiC$_x$ required *ex-situ* furnace post-processing.

Post-processing was performed according to the heating schedules and gas compositions presented in Figure 5 and Table 3. XRD results obtained on the converted cube surface and on the cube cross-section are shown in Figure 8. The TiC$_x$ yield obtained from cube surface characterization is reflective of the maximum carbide yield when gas-phase availability is not limited; phase characterization of the cube cross-section (in the x,z plane along the gas flow direction and perpendicular to the alumina substrate) is representative of the average chemical composition. Conversion results are summarized in Table 4.

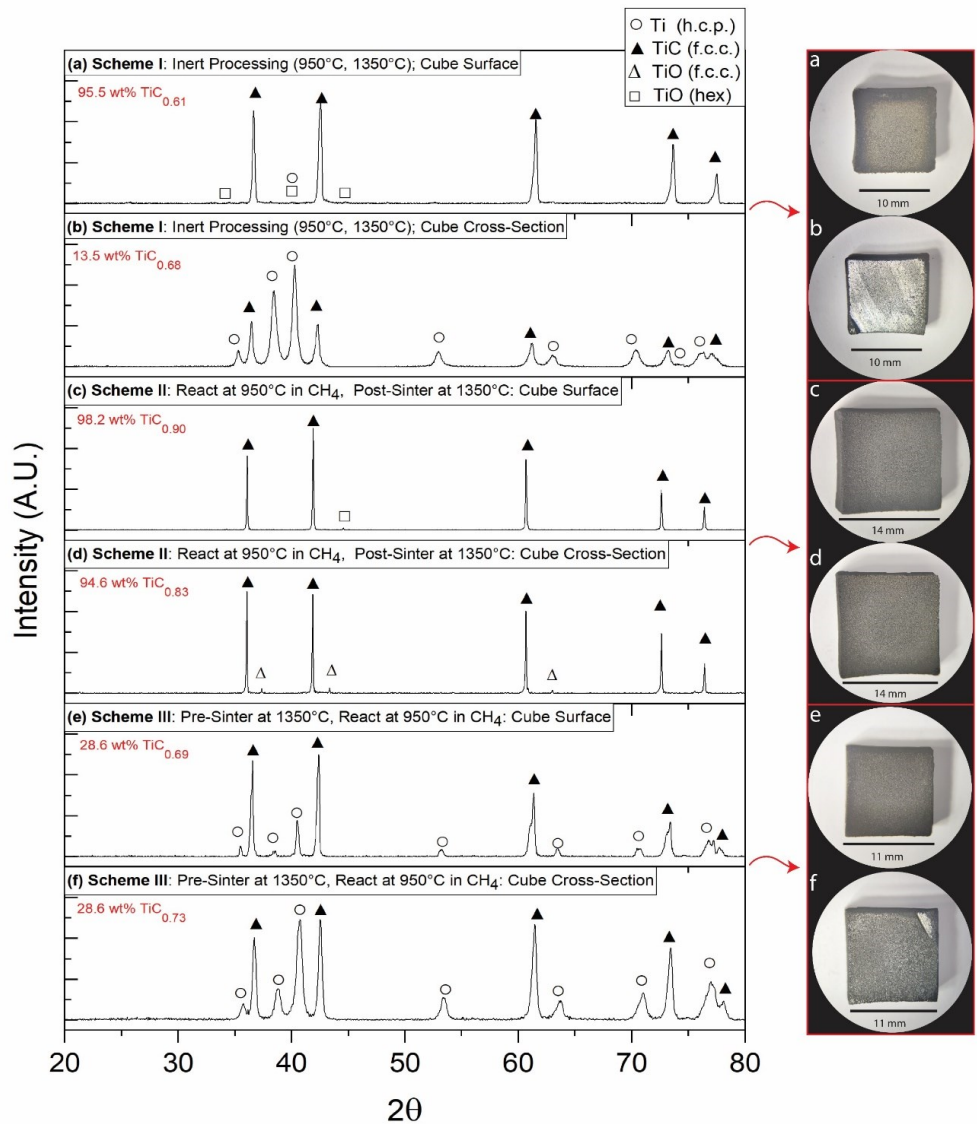

**Figure 8.** XRD spectra of post-processed Ti+phenolic parts converted to TiC$_x$. For each processing scheme, the optical images of the sample surfaces and cross-sections used for analysis are shown (a-f).

**Table 4. X-Ray Composition Analysis of Post-Processed Ti + Phenolic Cubes Using Rietveld Refinement Modeling**

| Characterization Area (Cube) | Characterization Area (Cube) | Total Titanium Carbide $TiC_x$ Product (wt%) | $TiC_x$, Lattice Parameter (Å) | Calc. NaCl-Type Product Stoichiometry | Unr. Ti (wt%) or C in α-Ti | Calc. Volume Change for conversion of Ti to $TiC_x$† | Notes |
|---|---|---|---|---|---|---|---|
| Scheme I: Inert Processing (Control) | Surface | 95.5% | 4.289 | $TiC_{0.61}$ | <1% | 11.30% | Trace TiO (hex) phase at 44.2° 2θ (<2.4 wt%) and TiO (cubic) |
| | Cross-Section | 13.5% | 4.300 | $TiC_{0.68}$ | 86.5% | 3.32% | Lack of reflection for β-Ti at 44.7° despite large peak at 38.4°. The lattice parameter of α-Ti: a=2.963 Å, c=4.717Å; suggests carbon incorporation in the lattice. |
| Scheme II: React, Post-Sinter | Surface | 98.2% | 4.322 | $TiC_{0.90}$ | 0% | 14.14% | Trace TiO (hex) phase at 44.2° 2θ (<1.8 wt%) |
| | Cross-Section | 94.6% | 4.317 | $TiC_{0.83}$ | 0% | 13.10% | TiO (fcc) phase at 37.2° amd 43.4° 2θ (<5.4 wt%) |
| Scheme III: Pre-Sinter, React | Surface | 81.5% | 4.301 | $TiC_{0.69}$ | 18.5% | 10.90% | Unknown trace impurity at 2.2° and 44.2° 2θ |
| | Cross-Section | 38.6% | 4.306 | $TiC_{0.73}$ | 61.4% | 6.47% | The lattice parameter of α-Ti: a=2.953 Å, c=4.710Å; suggests carbon incorporation in the lattice. |

† Product stoichiometry was estimated from lattice parameter methods described in Chapter 4 and [41][14].

Conversion at 950°C in 80/20 vol Ar/$CH_4$ resulted in up to 98.2 wt% $TiC_{0.90}$ yield from the 75 vol% Ti powder+ 25 vol% phenolic precursor mixture. Diffraction peaks at 36.0°, 41.8°, 60.6° 2θ in Figure 8 are indicative of NaCl-type $TiC_x$ ceramic. NaCl-type $TiC_x$ has a wide range of stoichiometries and interstitial carbon occupancies that range from $TiC_{0.47}$-$TiC_{1.0}$ [14], [23]. Using Rietveld refinement, the lattice parameters of the $TiC_x$ product phases were estimated to range from 4.289 - 4.322 Å. These values can be compared to the lattice parameter of 4.327 Å for stoichiometric $TiC_{1.0}$ [14],[41]. Quantitative assessment using empirically-derived, lattice parameter-chemistry relationships indicates that the product carbide stoichiometry varied between $TiC_{0.61}$ -$TiC_{0.90}$ and was highly dependent on processing parameters. XRD for both inert and reactive heating schemes do not show a residual carbon peak at 28 degrees 2θ (as compared to the green part in Figure 7). The results indicate that carbon associated with phenolic decomposition from SLS processing was consumed during tube furnace processing to TiC.

Using control Scheme I, significant surface conversion of the Ti+phenolic cube was achieved (95.5 wt%, $TiC_{0.61}$) without $CH_4$ gas reactivity. The carbide phase in the cube's cross-section was determined to be 13.5 wt% $TiC_{0.69}$ with a visually metallic inner core. In the absence of $CH_4$ gas-solid reactivity, carbonaceous compounds appeared to migrate to the exterior of the cube (and possibly exit the structure) before phenolic decomposition was complete. This was partially indicated by carbonized resin traces observed on the interior of the alumina tube after processing. By contrast, α-Ti and/or a solid solution of C and α-Ti was the dominant unreacted phase in the interior of the inert processed sample. If carbon from phenolic decomposition were completely consumed during the solid-state reaction, the estimated yield of $TiC_{0.61}$ would be approximately 51 wt%, assuming sample homogeneity. The conversion results indicate that the utilization of carbon supplied by phenolic binder was only ~26% efficient.

XRD analysis indicates that the addition of $CH_4$ to the post-processing atmosphere dramatically increased $TiC_x$ yield. The direct reaction of Ti and $C_{(s)}$ is thought to require higher temperatures than are needed for reactions with $CH_4$ which can rapidly occur at temperatures near 700 °C [42]. Post-processing of the Ti+ phenolic structures using Scheme II produced 98.2% surface $TiC_{0.90}$ and 95.1 wt% average $TiC_{0.83}$. No unreacted Ti precursor material was detected by XRD. TiO (at 37.2° and 43.3° 2θ in Figure 8) was the only other quantifiable trace component (<5.4 wt%). Oxygen contamination in the interior of the structure rather than on the top cube surface might be related to preferential oxidation of Ti particles by off-gassing phenolic decomposition products and more incomplete reduction in the interior of the sample with limited $CH_4$ gas-phase availability. Even so, results in Table 4 suggest that when structures were subject to gas-solid reactivity before high-temperature sintering at 1350 °C, the reaction was almost complete. The product composition, $TiC_{0.83}$, is very near the non-stoichiometric composition ($TiC_{0.78\pm0.03}$) with a maximum melting temperature of 3070 °C which far greater than the processing temperatures used [14].

In contrast to Scheme II, conversion via scheme Scheme III (pre-sintering followed by $CH_4$ reactivity) hindered gas-phase reactivity and appeared to prevent $CH_4$ penetration into the sintered particle mixture. The exterior of the sample was converted to 81.1 wt% $TiC_{0.69}$, while the interior sample composition was 38.6 wt% $TiC_{0.73}$ with α-Ti remaining as the primary unconverted phase. The melting temperature of Ti is approximately 1668 °C so initial heating at 1350°C significantly densifies the green body by thermal sintering – this occurs optimally between 2/3 - 3/4 $M_p$ or 1100-1250 °C for Ti [20]. The larger lattice parameter of α-Ti determined by Rietveld refinement (a=2.953Å, c= 4.671 Å, compared to 2.951 Å and = 4.686 Å) suggests solubility of carbon in the h.c.p. titanium lattice. The total integrated time during heating for the reactive processing schemes was identical at 54.8 hrs, however, the time of gas-solid reactivity within the overall processing timeline appears to dictate conversion efficiency and resultant volume change.

### 3.3 Volume Change and Morphological Assessment of Converted AM Structures

After de-binding, conversion, and consolidation, the AM-cube and lattice structures were measured to estimate the net volume changes associated with gas-solid conversion, densification and sintering. The dimension and mass/density changes of the samples are summarized in Table 5. Comparisons between the cube and lattice samples before and after furnace processing are shown in Figure 9.

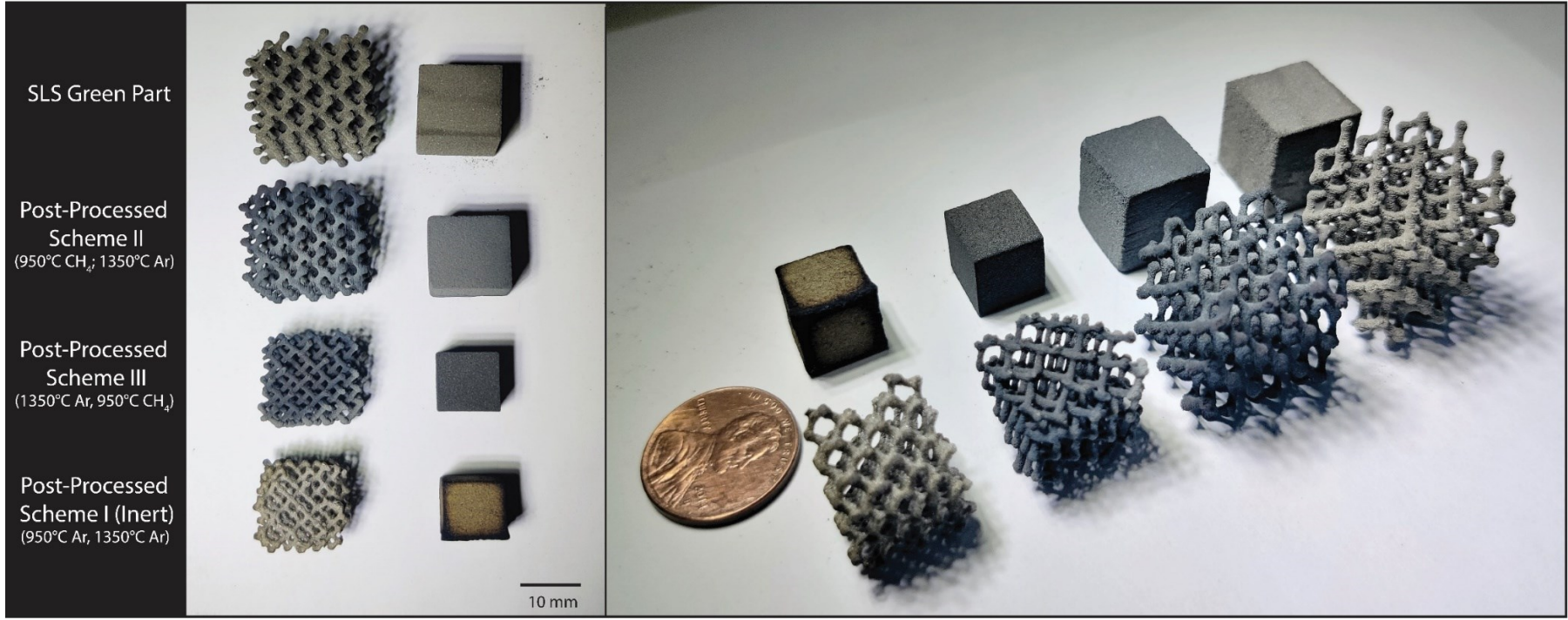

**Figure 9.** Photographs of the SLS green part in comparison to the post-processed samples converted to TiC$_x$ containing structures using Schemes I-III.

**Table 5. Summary of SLS Processed Cube Samples Pre- and Post-conversion in $CH_4$ to $TiC_x$**

| | Composition | X (mm) | Y (mm) | Z (mm) | Volume ($cm^3$) | Mass (g) | Density ($g/cm^3$) | Estimated Volume Fraction of Solid Phase |
|---|---|---|---|---|---|---|---|---|
| SLS Cube (Green) | α-Ti + Phenolic | 15.00 (SD=0.08) | 14.90 (SD=0.06) | 15.41 (SD=0.06) | 3.445 (SD=0.05) | 4.01 (SD=0.01) | 1.16 (SD=0.01) | 31.8% (SD=0.003) |
| Cube (Scheme I, Inert) | $TiC_{0.68}$ (13.5 wt%) α-Ti (86.5 wt%) | 10.50 - | 10.41 - | 10.43 - | 1.140 - | 3.91 - | 3.44 - | 73.8% - |
| *% Change, Inert* | - | -30.0% - | -30.1% - | -32.3% - | -66.9% - | -2.55% - | +197% - | +42.0% - |
| Cube (Scheme II: 950°C, 1350°C) | $TiC_{0.93}$ (94.6 wt%) | 14.17 (SD=0.07) | 14.00 (SD=0.06) | 14.36 (SD=0.03) | 2.847 (SD=0.007) | 4.34 (SD=0.005) | 1.52 (SD=0.002) | 31.8% (SD=0.01%) |
| *% Change* | - | -5.56% (SD=0.42%) | -6.04% (SD=0.41%) | -6.81% (SD=0.41%) | -17.32% (SD=0.22%) | +8.10% (SD=0.12%) | +30.7% (SD=1.21%) | 0.0% (SD=0.04%) |
| Cube (Scheme III: 1350°C, 950°C) | $TiC_{0.69}$ (38.6 wt%) α-Ti (61.4 wt%) | 10.90 (SD=0.02) | 10.72 (SD=0.01) | 10.96 (SD=0.02) | 1.281 (SD=0.01) | 3.91 (SD=0.005) | 3.06 (SD=0.001) | 52.3% (SD=0.04%) |
| *% Change* | - | -27.33% (SD=0.13%) | -28.05% (SD=0.07%) | -28.88% (SD=0.13%) | -68.82% (SD=0.17%) | -2.49% (SD=0%) | +162% (SD=1.20%) | -20.5% (SD=0.04%) |

*Green state values are from the five examples of the same part produced in a single build, while values the post-processed samples are the average of two sample conversions. The composition of converted samples are from the cross-sectional XRD. Standard deviations (SD) are indicated below each value.

Relatively uniform shrinkage of cube samples occurred across the x, y, and z from post-processing (Table 5). Increased consolidation in the z-direction was due to gravity. The estimated volume fraction of solid phase and porosity of the green samples were nearly identical using Scheme II, but not Scheme I (inert) or Scheme III (sinter, then react). When reactivity preceded conversion, the high melting point of the $TiC_x$ product phase (up to 3160 °C) prevented significant densification due to slow atomic diffusion. While stoichiometric $TiC_{1.0}$ was not achieved, a comparison of the results for samples processed in Scheme II and III suggests temperature control and heating duration can be used to alter green body microstructures and tailor conversion rates and carbide stoichiometry. This two-step post-processing procedure might prove most efficient in creating dense, and robust UHTC components if gas-solid reactivity is carried out before the green body is densified until gas diffusivity is limited. During this reaction synthesis process, temperature, gas composition and processing conditions must be carefully controlled to ensure simultaneous exothermic reactivity, reaction bonding, and densification to produce well-bonded, denser TiC parts. Through proper selection of post-processing times and temperatures, gas and carbon diffusion might be controlled to meet the length scales required for component features (e.g. thin lattice struts versus a dense cube).

Volume expansion from the conversion of Ti → $TiC_{1.0}$ used up to ~42% of the phenolic volume contained in the green part. This effect might be beneficial compared to non-reactive methods, but the high shrinkage inherent to multi-step AM processes cannot be wholly circumvented. Non-reactive SLS methods that incorporate 50-70 vol% organic binder materials are subject to anisotropic shrinkage (-36.8 vol% to -61.4 vol%), cracking, and low part densities ranging from 36-66% [38], [43]. These values are comparable to those presented in this work when reactivity was incomplete, and the brown body was α-Ti rich (i.e. 66.9% to 68.8% volume reduction). Using Scheme II, the combination of the chemically-induced volume changes and slow atomic diffusion for $TiC_x$ reduced shrinkage to -17.3% where interparticle bonding was achieved by gas-solid reactivity. In this case, the estimated volume fraction of solid phase of 31.8% was unchanged and identical to the green body. This indicates qualitatively that expansion from Ti to $TiC_x$ partially reduced overall consolidation due to phenolic burnout. Higher-temperature post-sintering (not accessible in this work) and or isostatic pressing could then be used after reactivity to increase density at the expense of some part shrinkage. This method (in comparison to direct densification of non-reactive particle-based UHTC green bodies [10]) might reduce defect structures generally observed for indirect UHTC AM. Samples using this two-step reactive approach were physically robust enough to be handled and were macroscopically crack-free as is shown in Figure 10.

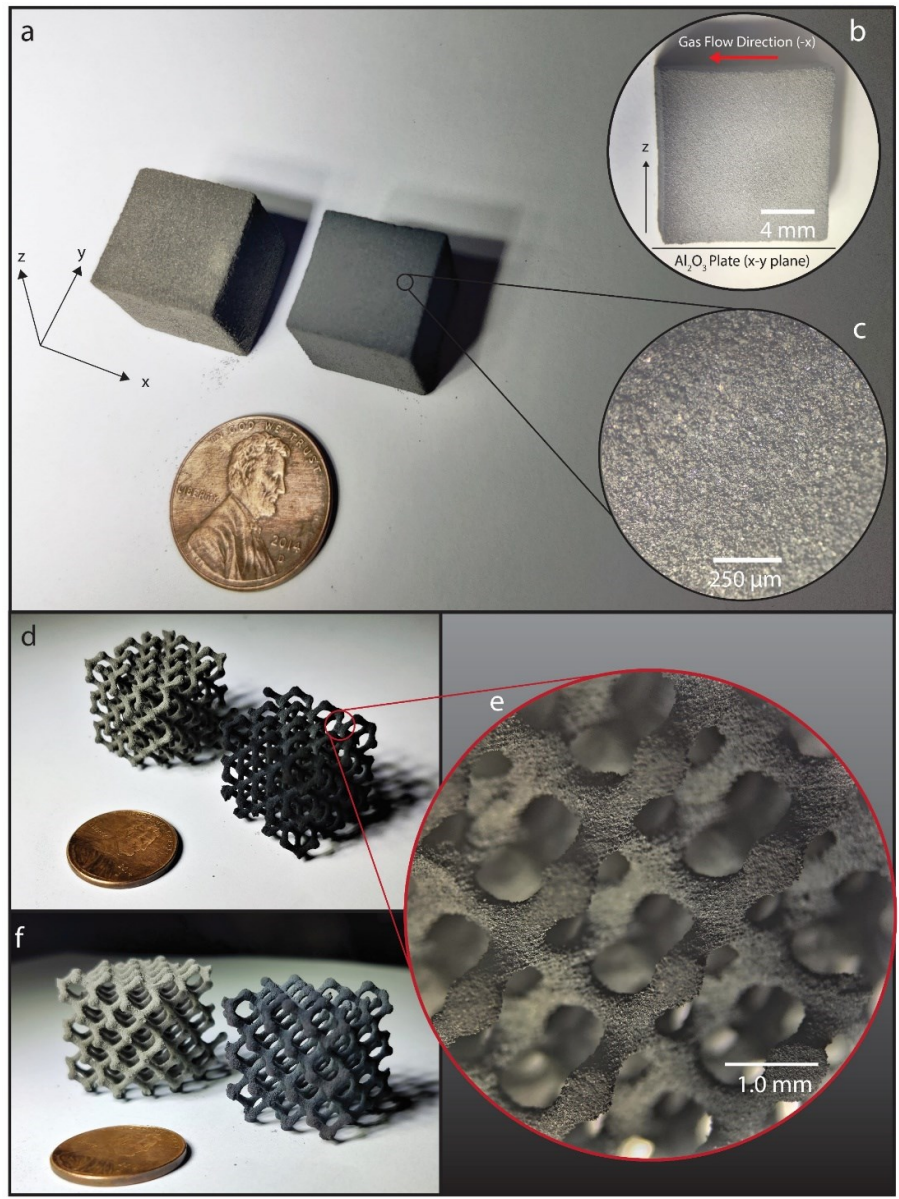

**Figure 10.** Photographs and photomicrographs depicting the SLS processed green Ti+phenolic cube samples before and after $CH_4$ post-processing to $TiC_x$. The leftmost structure in (a) is the green state cube. To the right of this sample is the $TiC_x$ cube after post-processing. (b) depicts the cross-section of the post-processed structure showing uniform conversion into the center of the structure under the prevailing reaction conditions. (c) shows the surface morphology of the $TiC_x$ cube. In (d) and (f) the leftmost sample is the unreacted green Ti+phenolic lattice, while the rightmost sample is the $TiC_x$ material

after post-processing. (e) is a high magnification image of the lattice morphology after CH$_4$ post-processing.

*3.3.1 Observations on the Volume Changes for Lattice Structures*

Similar shrinkage relationships were observed for lattice structures using Schemes I-III. The composition of these samples was not explicitly characterized via XRD; the stoichiometry and wt% fractions of TiC$_x$ are assumed to be greater than or equal to the cubes given their higher surface-area-to-volume ratios and shorter distances for diffusion. Figure 11 shows SEM images of the green and lattice structures processed using Schemes I-III. Several differences in the volume change response during brown body formation and subsequent sintering were observed:

- Close inspection of Figure 11 reveals that the cross-sectional diameter of struts disproportionately consolidated due to the high surface-area-to-volume and aspect ratios of the features compared to the overall lattice dimensions.
- Consolidation in the z-direction (compared to the x or y directions) was more significant with the smaller feature sizes of the lattice especially when the materials composition was α-Ti rich.
- Heating cycle variation influences the pore structure of the final product. For example, conversion of Ti to TiC before sintering using Scheme II, resulted in greater residual porosity and sharper features than Scheme III due to reduced atomic diffusivity of TiC compared to Ti at 1350°C.

These effects have been noted by others in non-reactive ceramics-AM processing schemes that require high-temperature sintering/consolidation [38], [43]. In future developments, anisotropic and geometry-dependent volume changes should be considered during the design of green parts to obtain the required final geometry.

**Table 6. Summary of SLS Processed Lattice Samples Pre- and Post-conversion in CH$_4$ to TiC.**

| Sample | Composition | X (mm) | Y (mm) | Z (mm) | Boundary Volume (cm$^3$) | Mass (g) |
|---|---|---|---|---|---|---|
| SLS Lattice (Green) | α-Ti + Phenolic | 22.06 (SD=0.13) | 22.05 (SD=0.20) | 22.72 (SD=0.08) | 11.05 (SD=0.14) | 1.58 (SD=0.02) |
| Lattice (Scheme I, Inert) | TiC$_x$ | 16.01 | 15.93 | 14.43 | 9.83 | 1.46 |
| *% Change, Inert* | | -27.4% | -28.6% | -36.5% | -67.2% | -6.3% |
| Cube (Scheme II: 950°C, 1350°C) | TiC$_x$ | 20.99 | 21.62 | 21.67 | 9.83 | 1.74 |
| *% Change* | | -4.9% | -2.0% | -4.6% | -11.0% | +9.78% |
| Cube (Scheme III: 1350°C, 950°C) | TiC$_x$ | 17.68 | 17.36 | 14.82 | 4.55 | 1.48 |
| *% Change* | | -19.9% | -22.3% | -34.8% | -60.2% | -4.4% |

*Values for the green state sample were averaged over five samples processed in the same build.

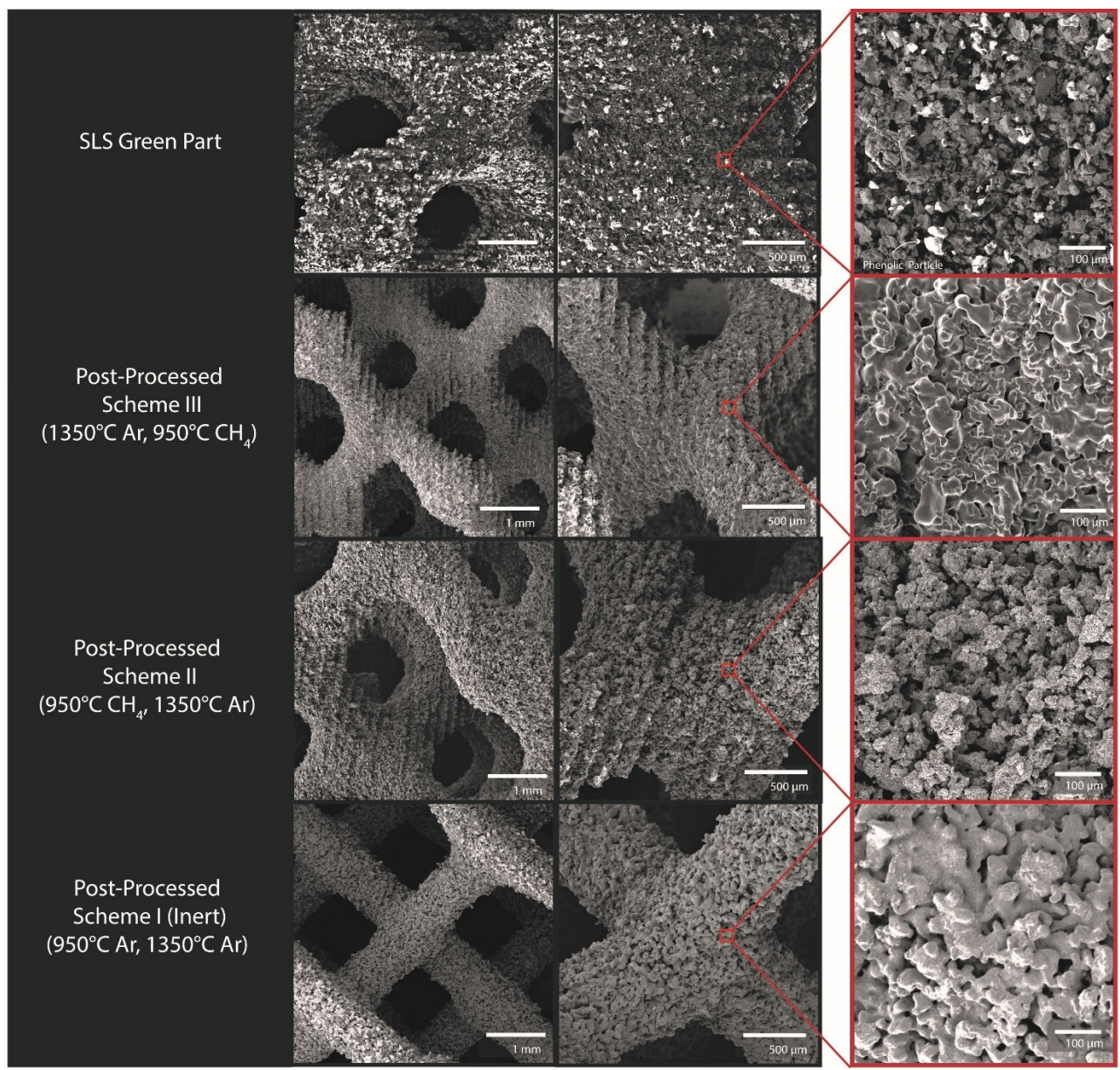

Figure 11. SEM images of lattice structures prior to post-processing (a) and after post-processing Schemes I-III (b-d). Samples are presented in order of descending macroscopic lattice size.

## 3.4 Gibbs Free Energy and TiC$_x$ Stoichiometry

Molecular dynamics simulations for C diffusivity in TiC$_x$ [30], revealed that as carbon stoichiometry increases, the activation energy required for diffusion significantly increases. The implication is that the processing conditions not only dictate the development of carbide phases but also affect the extent of reaction bonding. The exponential relationship relating NaCl-type TiC$_x$ stoichiometry (TiC$_{0.47}$-TiC$_{1.0}$) and activation energy for interlattice C diffusion is shown in Figure 12a. The required activation energies increase rapidly above TiC$_{0.9}$ and support experimental results where samples processed in CH$_4$ achieved a maximum interstitial occupancy of 0.83 (0.90 on the surface) after 12 hrs of reactive processing, after which significant energy requirements make stoichiometric conversion difficult to achieve.

Reaction thermodynamics are presented in Figure 12b where $\Delta G_r$ is plotted as a function of temperature for conversion of Ti by $C_s$ or $CH_4$. Thermodynamic calculations suggest that the exothermic reaction, $Ti + CH_4 \xrightarrow{950°C,} TiC + 2H_2$ is more thermodynamically favorable ($\Delta G°_r$=-215 kJ/mol) than the reactivity of Ti with $C_s$, $Ti + C_s \xrightarrow{950°C,} TiC$, ($\Delta G°_r$=-182 kJ/mol) after phenolic decomposition. By relating thermodynamic and kinetic data to processing conditions, differences in gas-solid versus solid-state reactivity are shown to impact the surface microstructure and propensity for particle bonding to occur. SEM image microscopy in Figure 12 illustrates that lattices subject to isothermal conversion in $CH_4$ before sintering at 1350 °C (Scheme II) had significant diffusion and/or reaction bonding that formed a continuous network of $TiC_x$. Due to the high melting point of $TiC_x$ ($M_p$~3000°C) relative to the post-processing temperature 1350 °C, sintering of the reacted structure can be largely excluded from the primary bonding mechanism. Here the $\Delta G°_r$ =-215 kJ/mol release may have facilitated rapid exothermic reaction propagation and reaction bonding through surface diffusion. The interparticle bonding observed from initial isothermal gas-solid conversion (even when conducted without heating at 1350 °C) was as significant as bonding induced from sintering of the Ti structure ($M_p$: 1668°C) when processed using identical but inert conditions (Scheme I). These results contrast the surface morphology of the structure obtained by post-processing using Scheme III, where particles appear discretely converted (Figure 12). Although reactivity-induced free energy release (in the experimental temperature range 25-1350 °C) cannot singularly overcome activation energies required for carbon diffusion through $TiC_{0.47}$-$TiC_{1.0}$ [14], [23], $\Delta G°_r$ can facilitate C diffusion through α-Ti ($E_a$=91 kJ/mol) or aid self-diffusion of Ti containing dilute carbon ($E_a$ =126 kJ/mol) [44]. Without isothermal gas-phase reactivity, carbide conversion is rate-limited by the decomposition of phenolic resin and conversion through slow solid-state carbothermal reactions [40]. This is likely to hinder development of conditions that enhance interparticle adhesion.

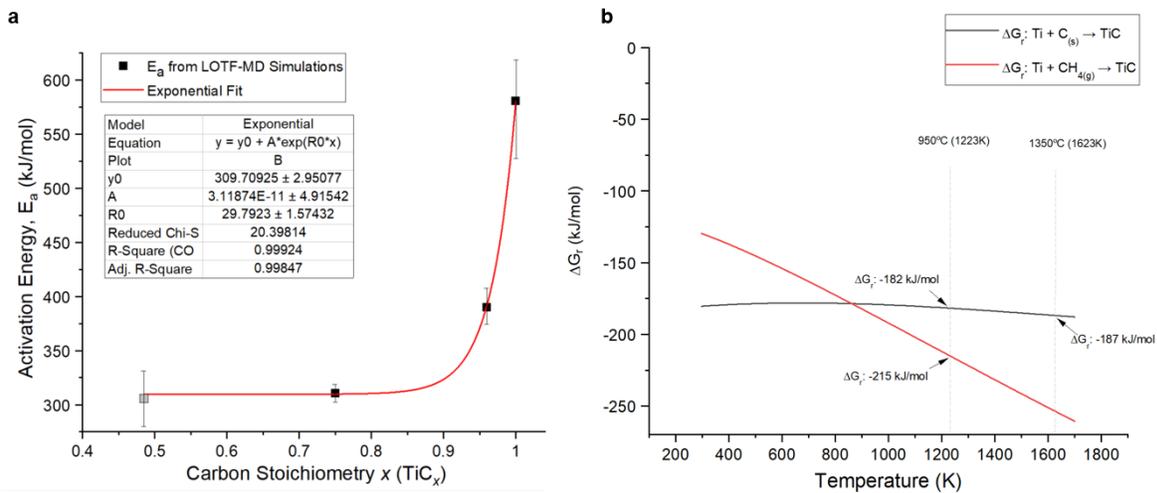

**Figure 12.** (a) The influence of carbon content on the activation energy required for C diffusion in $TiC_x$. The lower data point for $TiC_{0.48}$ was reported in [45]. (b) The temperature-dependent $\Delta G_r$ associated with Ti reaction with $C_s$ or $CH_4$.

Although a more detailed study on reaction bonding is required to fully understand the mechanisms involved, the additional exothermic energy relaese aids diffusion and appears to enhance interparticle bonding compared to other indirect AM techniques where gas-solid reactivity does not occur. For example, Deckers noted that extremely slow heating rates of ~6 °C/hr must be used for tube furnace de-binding of green bodies composed of non-reactive refractory ceramics particles, otherwise particle bonding would not occur [11]. Rates above 6 °C/hr with similar levels of porosity did not have any appreciable mechanical strength and readily crumbled [11]. In this work, a heating rate of 100 °C/hr was used without any significant structural defects. Even with slow heating rates, indirect AM of carbides, such as SiC, requires sintering and pyrolitic debinding of organic binders before reaction bonding to obtain appreciable mechanical properties [11]. In these multi-step methods, several phenolic impregnations, pyrolysis, and silicon infiltration cycles may be coupled with isostatic pressing to increase part densities beyond 90%, however, residual Si (or graphite) is often present - typically 10-15 wt% [46], [47]. In future efforts, TiC part densities might be increased through hot isostatic pressing [47], or molten infiltration of metals like Al or Si to form TiC-Al cermets [48]or $Ti_3SiC_2$ MAX phase compositions respectively [49]. Even so, samples fabricated in preliminary trials that relied solely on 950 °C reaction synthesis (8 hrs dwell, no 1350 °C sintering) were robust enough to be easily handled, macroscopically defect-free, and nearly fully converted to carbide. Without reactivity, the higher melting point of TiC (compared to SiC) might prevent robust UHTC parts from being obtained using standard indirect SLS methodologies when processing temperatures are well below what is ordinarily required for UHTC sintering (T > 2000 °C).

## 3.5 Qualitative Thermal Shock Testing of $TiC_x$ Lattice

Lattice structures were thermally stressed using continuous propane torch heating to demonstrate their refectory characteristics and resistance to thermal shock. The lattice produced using Scheme II was used for testing because it contained the greatest phase fraction $TiC_x$ and the highest theoretical melting temperature due to its $\sim TiC_{>0.83}$ stoichiometry. This sample was positioned approximately 25 mm from the end of the propane torch. Once lit, the lattice was subject to 120 seconds of continuous heating using the hottest portion of the blue inner flame cone. The air-fed propane torch was estimated to produce flame temperatures of approximately 1300 °C [50]. A video demonstration of high-temperature testing is shown in Figure 13.

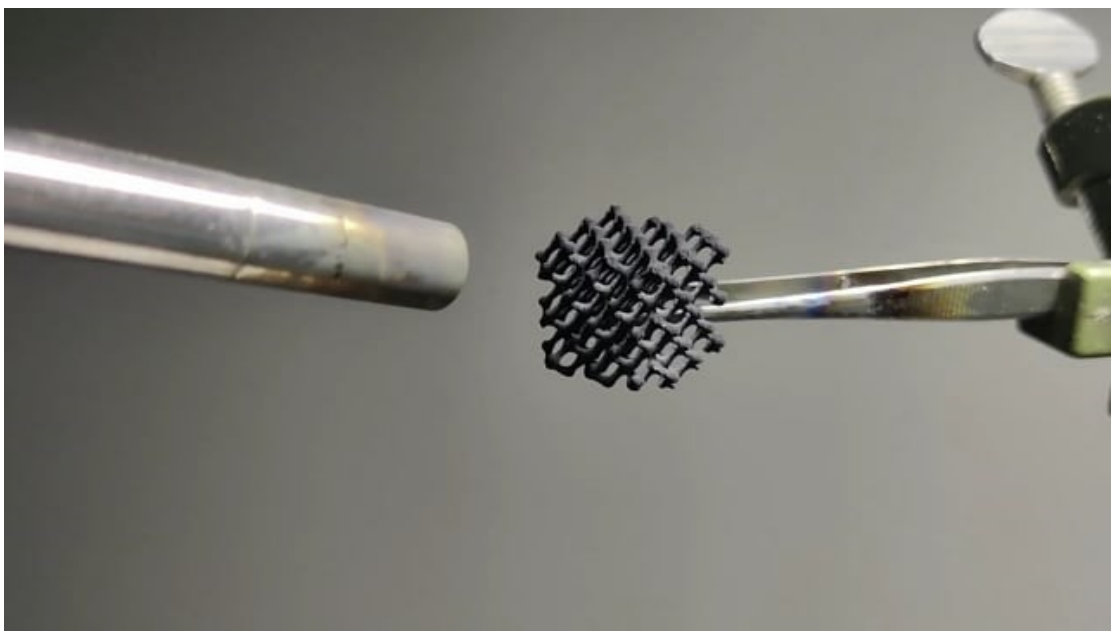

**Figure 13.** Video demonstration of TiC$_x$ lattice (fabricated by post-processing using Scheme II) that was heated using a propane torch.

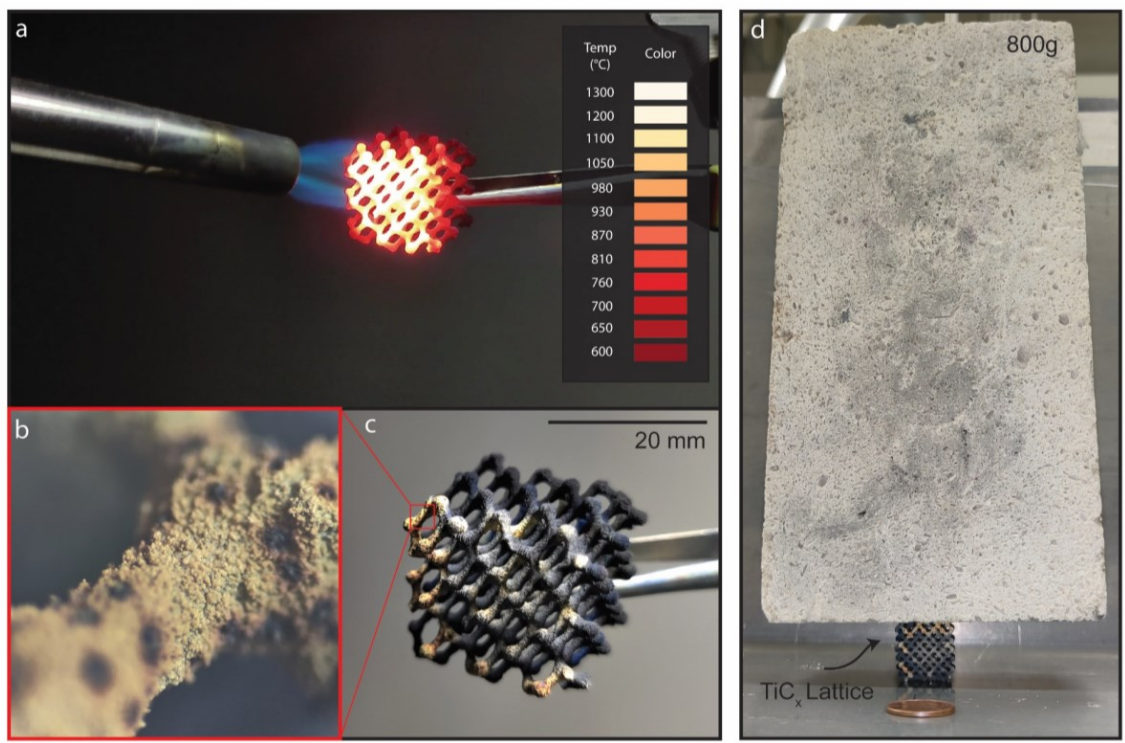

**Figure 14.** Photograph still from the recorded blow torch test (a) and optical micrograph of resulting microstructure (b), and photographs of the product lattice (c). (d) shows the lattice after heating supporting an 800 g alumina firebrick to illustrate its qualitative mechanical properties; a penny is shown for scale.

The flame-facing surface of the lattice in Figure 14 reached the torch's peak temperature according to a qualitative estimate based on its temperature dependant, thermal emission/radiation (also known as Blackbody radiation) [51]. A digital pyrometer was initially used but it reached its maximum operational reading of 1190 °C and was unable to provide temperatures beyond this upper limit. Radiative glow temperature estimation suggests that the front of the lattice reached a peak temperature of 1300 °C within 3.5 seconds of heating. Despite the small dimensions of the lattice, the back half of the structure reached thermal equilibrium at ~600 °C, while still being enveloped in the flame. The lattice did not experience a mechanical failure or cracking due to $TiC_x$'s low coefficient of thermal expansion (CTE, $5.2$-$7.4 \times 10^{-6}$ $K^{-1}$ [14]), residual porosity, and reaction bonded structure [6]. As the material was heated, high heat capacity (50.5 J $mol^{-1}$ $K^{-1}$ at 25°C, 53.8 J $mol^{-1}$ $K^{-1}$ at 1000°C [14]) and low thermal conductivity (17.4 W $m^{-1}$ $K^{-1}$ at 25°C to 5.24 W/m/k at 1000°C which declines with carbon deficiency [14], [52]) prevents heat transfer before the heat is promptly released due to high emissivity (~ 0.9 [53]). The effects of high surface area are improved by the macroscopic and microscopic porosity the $TiC_{0.83}$ structure which is only 3.7% of the occupied lattice volume. This combination of physicochemical properties is similar to the highly porous (90% void, 10% $SiO_2$) HSRI Space Shuttle Tiles used for thermal protection systems [54].

In Figure 14 b-d oxidation to $TiO_xC_{1-x}$, $TiO_x$, and $TiO_{2-x}$ was apparent on the extremities of the lattice structure. Oxidation of $TiC_x$ to TiO, $TiO_{2-x}$ (rutile) and $TiO_{2-x}$ (anatase) all form protective scales with Pillings-Bedworth ratio (the ratio of the oxide volume to the material volume, TiC) of $R_{PB}$=1.06, $R_{PB}$=1.55, and $R_{PB}$=1.68, respectively that might fill in residual porosity and slow further reactivity. Even after thermal shock testing, Figure 14d indicates the AM lattice structure qualitatively maintained useful mechanical properties and was able to support an 800 g alumina firebrick. The culmination of these properties makes AM structures produced in this work of interest for further investigation including mechanical and thermal testing of useful refractory UHTC components.

## 3.6 Conclusions

In this work, a two-step reactive AM approach was studied for the formation of the ultra-high temperature ceramic $TiC_x$. Readily available equipment including a polymer powder bed fusion AM machine and traditional tube furnace were used to produce complex UHTC parts. This processing scheme incorporated, (1) selective laser sintering of Ti precursor mixed with a phenolic binder for green body shaping, and (2) *ex-situ,* isothermal gas-solid conversion of the green body in $CH_4$ to form a $TiC_x$ part. Three different heating schedules were investigated for efficient reactivity and volume control of the green 15 mm × 15 mm × 15 mm cubes (Scheme I: inert gas processing; Scheme II: gas-solid reactivity then post-sintering; Scheme III: pre-sintering then gas-solid reactivity). The results indicate that phenolic decomposition during post-processing facilitated a cross-sectional $TiC_{0.68}$ yield of 13.5 wt%. Meanwhile, the reactive processing in $CH_4$ was effective in promoting conversion to $TiC_x$ of varied carbide stoichiometries. A maximum yield of 98.2 wt% $TiC_{0.90}$ (94.6 wt% $TiC_{0.83}$ interior) was achieved when samples were first converted in 80/20 Ar/$CH_4$ at 950 °C then sintered at 1350 °C in Ar. When heating and reactivity were carried out in reverse order, higher volume fraction of solids but more shrinkage (-68.8% vs -17.3%) were induced due to pre-sintering of Ti before UHTC formation. With the complete conversion of Ti to $TiC_x$, molar volume expansion (+14.2%) appears to partially compensate for phenolic binder decomposition and the densification process for reduced component shrinkage as compared to non-reactive indirect SLS processing. Prevailing reaction kinetics (carbon diffusion through $TiC_x$) and Gibbs free energy release appear to control the product stoichiometry and reaction bonding behavior that enables UHTC particles to bond during low-temperature processing that is otherwise unachievable.

Complex lattice geometries were also fabricated using this AM methodology to investigate the reactive AM approach for producing intricate structures with small features (<800 μm struts, ~50 μm resolution). For all post-processing schemes, resulting lattices were robust enough to be easily handled and crack-free. The TiC$_{\geq 0.83}$ lattice produced using Scheme II was subjected to rapid, high-temperature heating to characterize material response to extreme thermal loads. The unique combination of TiC$_{\geq 0.83}$ materials properties and the complex AM structure allowed the lattice to reach a peak steady-state temperature of 1300 °C for 2 minutes with minimal oxidation and without fracture. If denser UHTC components are desired, then higher-temperature post-sintering (not accessible in this work) and or isostatic pressing could be used after conversion to increase density. Broadly, with additional development and investigation, this additive manufacturing approach appears could be viable for the production of UHTC carbides such as ZrC, HfC, or TaC that are otherwise incompatible with prevailing AM techniques that do not incorporate reaction synthesis techniques.

## 4. Declaration of Competing Interest

Funding for the study described in this publication was provided by the Office of Naval Research. Under a license agreement between Synteris LLC and the Johns Hopkins University, Dr. Peters and the University are entitled to royalty distributions related to the technology described in the study discussed in this publication. Dr. Peters is a founder of and holds equity in Synteris LLC. He also serves as the Chief Technical Officer and holds a board seat at Synteris LLC. The results of the study discussed in this publication could affect the value of Synteris. This arrangement is pending review and approval by the Johns Hopkins University in accordance with its conflict of interest policies. Dennis Nagle, and Dajie Zhang are entitled to royalty distributions related to the technology described in the study discussed in this publication. Dr. Nagle, and Dr. Zhang are not affiliated with Synteris LLC. This arrangement is approved by the Johns Hopkins University in accordance with its conflict of interest policies.

## 5. Acknowledgments

The authors gratefully acknowledge funding provided by the Office of Naval Research, Nanomaterials Program Office, under contract N00014-16-1-2460 in partial support of this research. We would also like to thank the Johns Hopkins Applied Physics Laboratory Graduate Fellowship Committee for graduate student support provided for A.B. Peters.

**Reference**

[1]   I. Gibson and D. Rosen, *Additive Manufacturing Technologies*, Second. Springer, 2015.

[2]   T. D. Ngo, A. Kashani, G. Imbalzano, K. T. Q. Nguyen, and D. Hui, "Additive manufacturing (3D printing): A review of materials, methods, applications and challenges," *Compos B Eng*, vol. 143, no. December 2017, pp. 172–196, 2018, doi: 10.1016/j.compositesb.2018.02.012.

[3]   A. Zocca, P. Colombo, C. M. Gomes, and J. Günster, "Additive Manufacturing of Ceramics: Issues, Potentialities, and Opportunities," *Journal of the American Ceramic Society*, vol. 98, no. 7, pp. 1983–2001, 2015, doi: 10.1111/jace.13700.

[4]   W. Fahrenholtz, Wuchina. E, W. Lee, and Y. Zhou, *Ultra High Temperature Ceramics Materials for Extreme Environment Applications*. 2014.

[5]   T. A. Parthasarathy, M. D. Petry, M. K. Cinibulk, T. Mathur, and M. R. Gruber, "Thermal and oxidation response of UHTC leading edge samples exposed to simulated hypersonic flight conditions," *Journal of the American Ceramic Society*, vol. 96, no. 3, pp. 907–915, 2013, doi: 10.1111/jace.12180.

[6]   T. H. Squire and J. Marschall, "Material property requirements for analysis and design of UHTC components in hypersonic applications," *J Eur Ceram Soc*, vol. 30, no. 11, pp. 2239–2251, 2010, doi: 10.1016/j.jeurceramsoc.2010.01.026.

[7]   E. Feilden, D. Glymond, E. Saiz, and L. Vandeperre, "High temperature strength of an ultra high temperature ceramic produced by additive manufacturing," *Ceram Int*, vol. 45, no. 15, pp. 18210–18214, 2019, doi: 10.1016/j.ceramint.2019.05.032.

[8]   W. G. Fahrenholtz and G. E. Hilmas, "Ultra-high temperature ceramics: Materials for extreme environments," *Scr Mater*, vol. 129, pp. 94–99, 2017, doi: 10.1016/j.scriptamat.2016.10.018.

[9]   T. Chartier, "Ceramic Forming Processes," in *Ceramic Materials*, P. Boch and J.-C. N. Niepce, Eds. London, UK: ISTE, 2007, pp. 123–197. doi: 10.1002/9780470612415.ch5.

[10]  M. C. Leu, S. Pattnaik, and G. E. Hilmas, "Investigation of laser sintering for freeform fabrication of zirconium diboride parts," *Virtual Phys Prototyp*, vol. 7, no. 1, pp. 25–36, Mar. 2012, doi: 10.1080/17452759.2012.666119.

[11]  S. Meyers, "Additive Manufacturing of Technical Ceramics Laser Sintering of Alumina and Silicon Carbide," KU Leuven, 2019.

[12]  L. Toth, *Transition Metal Carbides and Nitrides*. Academic Press, 1971.

[13]  Y. Suda, H. Kawasaki, K. Doi, J. Nanba, and T. Ohshima, "Preparation of crystalline TiC thin films grown by pulsed Nd:YAG laser deposition using Ti target in methane gas," *Mater Charact*, vol. 48, no. 2–3, pp. 221–228, 2002, doi: 10.1016/S1044-5803(02)00243-7.

[14]  I. L. Shabalin, "Titanium Monocarbide," in *Ultra-High Temperature Materials III*, Dordrecht: Springer Netherlands, 2020, pp. 11–514. doi: 10.1007/978-94-024-2039-5_2.
27